# MULTISCALE MODELING OF PLASTIC DEFORMATION OF MOLYBDENUM AND TUNGSTEN: III. EFFECTS OF TEMPERATURE AND PLASTIC STRAIN RATE

R. Gröger[1,2]* and V. Vitek[1]

[1] University of Pennsylvania, Department of Materials Science and Engineering,
Philadelphia, PA 19104, USA
[2] Los Alamos National Laboratory, Theoretical Division, Los Alamos, NM 87545, USA

*Corresponding author. *E-mail:* groger@lanl.gov

**Abstract**
In this paper we develop a link between the atomic-level modeling of the glide of 1/2⟨111⟩ screw dislocations at 0 K and the thermally activated motion of these dislocations via nucleation of pairs of kinks. For this purpose, we introduce the concept of a hypothetical Peierls barrier, which reproduces all the aspects of the dislocation glide at 0 K resulting from the complex response to non-glide stresses and expressed in a compact form by the yield criteria advanced in Part II. To achieve this the barrier is dependent not only on the crystal symmetry and interatomic bonding but also on the applied stress tensor. Standard models are then employed to evaluate the activation enthalpy of kink-pairs formation, which is now also a function of the full applied stress tensor. The transition states theory links then this mechanism with the temperature and strain rate dependence of the yield stress.

## 1. Introduction

In the Part I of this series we presented atomic-level studies of the glide of straight 1/2⟨111⟩ screw dislocations in Mo and W and in the Part II we used results of these calculations to analyze deformation behavior of single crystals. Since the atomistic calculations employed molecular statics they did not include any effects of temperature. Hence, both these studies deal with dislocation glide at 0 K although, as demonstrated in the Part II by comparison with experiments, important characteristics of the plastic yielding at low temperatures are well accounted for in this framework. Owing to the non-planar structure of the cores of 1/2⟨111⟩ screw dislocations the intrinsic lattice resistance to their glide is high and thus the applied shear stress needed for their motion at 0 K, customarily called the Peierls stress, is very high when compared, for example, with that of dislocations in FCC metals [3-8]. The models of the intrinsic lattice resistance to the dislocation glide all invoke the concept of the Peierls barrier that the dislocation has to overcome [9-11]. The Peierls stress is then the stress necessary to surmount this barrier at 0 K.

At finite temperatures the Peierls barrier can be surpassed with the aid of thermal activation via nucleation of pairs of kinks that subsequently migrate easily along the dislocation line. This process results in the dislocation glide in the direction of the kink formation [11-13]. Following the standard transition state theory of thermally activated processes (see for example [14, 15]) the velocity of the dislocation is then

$$v_d = v_0 \exp\left[-\frac{H(\boldsymbol{\sigma})}{k_B T}\right] \quad (1a)$$

and the plastic strain rate associated with the corresponding slip system is

$$\dot{\gamma} = \dot{\gamma}_0 \exp\left[-\frac{H(\boldsymbol{\sigma})}{k_B T}\right]. \tag{1b}$$

Here is the activation enthalpy that is a function of the applied stress tensor $\boldsymbol{\sigma}$, $k_B$ the Boltzmann constant, $T$ absolute temperature and $v_0$ and $\dot{\gamma}_0$ are the pre-exponential factors depending on the details of the mechanism of kink-pair formation that vary only weakly with the applied stress [11-15]. The yield stress, $\sigma_Y$, identified for a given slip system with the shear component of the applied stress acting in the slip plane parallel to the slip direction, can then be calculated from equation (1b) as a function of temperature and strain rate. At this point it has to be recognized that $\sigma_Y$ is composed of two terms:

$$\sigma_Y(T,\dot{\gamma}) = \sigma^*(T,\dot{\gamma}) + \bar{\sigma}, \tag{2}$$

where $\sigma^*$ is the temperature and strain rate dependent component of the yield stress and $\bar{\sigma}$ is the athermal stress that $\sigma_Y$ approaches at high temperatures [12, 15]. The origin of $\bar{\sigma}$ is any long-range stress field that opposes the dislocation glide, such as that associated with other dislocations. Since $\sigma^*$ is the shear stress in the slip plane parallel to the slip direction, only $\sigma^*$ does work during the activation process, but other constituents of the activation enthalpy may depend on the full stress tensor. This model has been very successful in theoretical treatments of a number of dislocation phenomena involving the lattice resistance to the dislocation glide [13], such as internal friction (Bordoni peak) [1, 16, 17], thermally activated glide in semiconductors where the high lattice friction arises owing to the covalent bonds [18-20] and thermally activated glide of screw dislocations in BCC metals [2, 21, 22].

The models of the formation of kink-pairs assume implicitly that the Peierls barrier can be identified with the potential energy that varies periodically with the position of the dislocation in its slip plane. At 0 K the dislocation is then assumed to move under the effect of an applied stress continually through the lattice, attaining gradually metastable positions with higher energy. When the applied stress is equal to the Peierls stress no metastable position exists and the motion becomes perpetual. This picture of the dislocation glide is, indeed, appropriate when the dislocation core is planar, spread in the glide plane. However, in the case of screw dislocations in BCC metals, and generally whenever the reason for the high Peierls stress is the non-planarity of the core, the dislocation does not move progressively through the lattice. Rather, the core structure gradually transforms under the influence of the applied loading and the Peierls stress is the stress at which the core structure is sufficiently modified for the dislocation to start moving. This process has been observed in many molecular statics studies of the glide of $1/2\langle 111\rangle$ screw dislocations in BCC crystals at 0 K (see, for example, [5, 23-26] and Part I). However, while the Peierls stress is readily obtained in such calculations, the corresponding Peierls barrier cannot be directly ascertained.

A conceivable direct atomistic approach to study the glide of $1/2\langle 111\rangle$ screw dislocations at finite temperatures is molecular dynamics simulation of the dislocation motion. Such calculations have been recently performed using a variety of central-force potentials of the Embedded Atom Method type [27-30]. These calculations, indeed, demonstrated that at applied stresses lower than the Peierls stress the motion of the screw dislocation proceeds via formation of pairs of kinks. However, the problem common to all molecular dynamics studies is the difficulty to capture rare events [31], such as formation of kink-pairs, the frequency of which is many orders of magnitude lower than the vibrational frequency of atoms. In Refs. [27-30] this was achieved by applying stresses very close to the Peierls stress but in this case the dislocation velocity and related strain rate are orders of magnitude larger than in the usual deformation studies. This problem has been discussed in detail in Ref. [29]. The molecular dynamics studies of the dislocation motion at much lower stresses, which should also encompass detailed investigation of the effects of shear stresses other than parallel to the slip direction, are not feasible at present.

An alternative approach is to investigate the activation path connecting the two subsequent positions of the screw dislocation by using the Nudged Elastic Band method for determining the saddle-point activated states [32, 33] (see also Appendix). This procedure was adopted in [34, 35] and kink-pairs were observed to form along these activation paths. The activation enthalpy for this process was then determined as a function of the applied shear stress parallel to the Burgers vector. Using this procedure it would be possible, at least in principle, to determine the dependence of the activation enthalpy on all the components of the stress tensor that were identified in Part I as affecting the Peierls stress. Notwithstanding, in this paper we propose a simpler approach that utilizes the results of 0 K calculations of the Peierls stress.

While the molecular statics calculations at 0 K do not determine the Peierls barrier, we develop a phenomenological paragon in which such barrier is regarded as a periodically varying function of the position of the dislocation that possesses all the features resulting from the non-planarity of the dislocation cores and its stress-induced transformations, in particular it reflects the dependence of the Peierls stress on both the sense of shearing in the direction of the Burgers vector and shear stresses perpendicular to the Burgers vector. The thermally activated process of the formation of kink-pairs can then be treated using the dislocation models advanced earlier [1, 2]. These models are briefly introduced in Section 2. Section 3 is then devoted to the most important development of this paper, construction of the Peierls potential and Peierls barrier on the basis of data attained in molecular statics studies presented in Part I. In contrast with previous treatments, this barrier, similarly as the Peierls stress, depends on the applied stress tensor, in particular shear stresses parallel and perpendicular to the slip direction in two {110} planes of the zone of the Burgers vector. Using the constructed Peierls barrier the activation enthalpy for the formation of kink-pairs is determined as a function of the applied stress tensor and corresponding temperature and strain rate dependence of the yield stress is evaluated in Section 4. Comparison of the calculated temperature dependence of the yield stress with available experimental data for Mo and W is presented in Section 5. This comparison demonstrates that the model of the formation of kink-pairs, enhanced by incorporation of the stress dependencies discovered in atomistic calculations, determines accurately not only the temperature and strain rate dependence of the yield stress but also reflects correctly the tension-compression asymmetry that arises from both the twinning-antitwinning asymmetry of shearing parallel to the slip direction and the effect of shear stresses perpendicular to the slip direction.

## 2. Mesoscopic dislocation models of kink-pair formation

The dislocation models of the formation of kink-pairs, developed originally in [1, 2, 16, 21], all assume that the dislocation glides in a well-defined slip plane and the Peierls barrier is a periodic function of one variable, the coordinate perpendicular to the dislocation line that is the direction of the dislocation glide. Hence, the terms Peierls potential and Peierls barrier have the same $V(\xi)$ meaning and are interchangeable. However, $1/2\langle 111\rangle$ screw dislocations in BCC metals do not have unique slip planes and, moreover, the Peierls barrier actually relates to the core transformations. Hence, we follow the suggestion of Edagawa et al. [36] who introduced the notion of the Peierls potential, $V(x,y)$, that is a function of two variables, $x$ and $y$, which represent the position of the intersection of the dislocation line with the {111} plane perpendicular to the corresponding $\langle 111\rangle$ slip direction. The transition of the screw dislocation between two low-energy sites at 0 K is then regarded as a motion along a minimum energy path, described by a curvilinear coordinate $\xi$, which mimics the core transformation, and the Peierls barrier, $V(\xi)$, is identified with the profile of the two-dimensional Peierls potential along the path $\xi$. In this representation both the Peierls potential and the Peierls barrier are regarded as dependent on the applied stress tensor. The details of the construction of the Peierls potential and the corresponding Peierls barrier are presented in Section 3.

In the following description of the mechanisms of kink-pair formation, employed when evaluating the corresponding activation enthalpies, we assume that the Peierls barrier, $V(\xi)$, is known and varies with the position of the dislocation along an activation path designated by $\xi$. The movement of the dislocation along $\xi$ proceeds by formation and subsequent extension of the kink-pairs. However, before such a kink-pair is formed segments of the dislocation vibrate and bow-out to various intermediate configurations. In most cases they return back to the original straight position. The enthalpy associated with these configurations, which includes the work done by the shear stress $\sigma^*$ that drives the dislocation glide, first increases with the bow-out but it reaches a maximum at a saddle-point for which the configuration is somewhere between the original straight line and the fully formed pair of well-separated kinks.

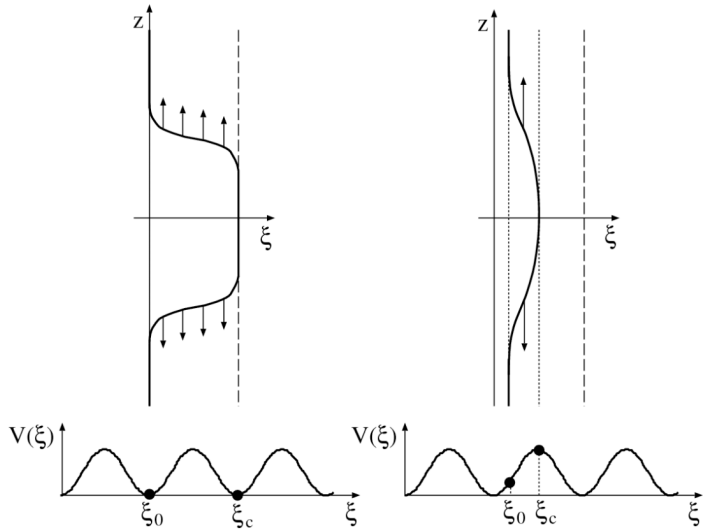

Fig. 1: Schematic illustration of the two types of saddle-point configurations associated with nucleation of a pair of kinks on the screw dislocation: (a) pair of well-developed kinks at low stresses and high temperatures, (b) the critical bow-out at high stresses and low temperatures. $\xi$ is a curvilinear coordinate along the activation path and the Peierls barrier along this path.

Two possible saddle-point configurations arise. The maximum enthalpy is attained either only after the pair of kinks has formed, as shown schematically in Fig. 1a, or when there is just a continuous bulge on the dislocation, as shown schematically in Fig. 1b. The former must take place at stresses approaching zero and as the stress increases the transition to the latter occurs. In the low-stress regime the fully developed kinks interact elastically via the attractive Eshelby potential that is in the framework of the isotropic elasticity equal to $-\mu a_0^2 b^2 / 8\pi\Delta z$ [13, 15], where $\mu$ is the shear modulus, $b$ the magnitude of the Burgers vector, $a_0$ the height of the kinks or, equivalently, the distance between two neighboring Peierls valleys in the slip plane, and $\Delta z$ their separation. The applicability of the elastic treatment of the kink-pairs was demonstrated in atomistic studies of Duesbery [37, 38]. This attractive interaction is opposed by the component $\sigma^*$ of the applied shear stress that drives the dislocation glide and does work $\sigma^* a_0 b \Delta z$ during the nucleation of the kink-pair. Such saddle-point configuration was first treated theoretically by Seeger [1, 16] in the context of studying internal friction. The enthalpy associated with this configuration is

$$2H_k - \frac{\mu a_0^2 b^2}{8\pi\Delta z} - \sigma^* a_0 b \Delta z, \tag{3a}$$

where $2H_k$ is the energy of two isolated kinks at zero stress and the last term is the work produced by $\sigma^*$ during the nucleation of the kink-pair. The separation of kinks for which the enthalpy (3a) attains a maximum defines the saddle-point configuration and the corresponding activation enthalpy of kink-pair formation is

$$H_{kp} = 2H_k - (a_0 b)^{3/2} \sqrt{\frac{\mu\sigma^*}{2\pi}}. \tag{3b}$$

In the high-stress regime we first have to recognize that the dislocation is moved away from its position in the unstressed crystal along the reaction coordinate to a position $\xi_0$ determined by the condition

$$\sigma^* b = \left.\frac{dV(\xi)}{d\xi}\right|_{\xi=\xi_0}. \tag{4a}$$

At this position the force originating from the Peierls barrier is equal to the Peach-Koehler force $\sigma^* b$. Owing to thermal vibrations the dislocation bows out towards the top of the Peierls barrier, as shown in Fig. 1b. If the bowed segment is described by a function $\xi(z)$, where the coordinate $z$ is in the direction of the originally straight dislocation, the energy associated with the bowed dislocation is

$$E_b = \int_{-\infty}^{+\infty} \left\{ [V(\xi) + E]\sqrt{1+\xi'^2} - [V(\xi_0) + E] \right\} dz, \tag{4b}$$

where $E$ is the line tension of the straight dislocation. The work done during the bow-out is

$$W = \int_{-\infty}^{+\infty} \sigma^* b(\xi - \xi_0) dz. \tag{4c}$$

The enthalpy associated with the bowing dislocation is $E_b - W$. Following Dorn and Rajnak [2] $\xi(z)$ is determined by functional minimization of this enthalpy with respect to $\xi$ and the corresponding Euler-Lagrange equation determines $d\xi/dz$ ($=\xi'$) as a function of $z$. The saddle-point configuration, corresponding to a critical value of $\xi = \xi_c$, is then determined by the requirement that the slope of the dislocation line is zero for $z = 0$. After reaching this configuration the dislocation line continues extending as a pair of kinks. This leads to the following condition determining $\xi_c$:

$$V(\xi_c) = \sigma^* b(\xi_c - \xi_0) + V(\xi_0). \tag{5}$$

The enthalpy associated with this configuration, i.e. the activation enthalpy of the kink-pair formation, is

$$H_b = 2\int_{\xi_0}^{\xi_c} \sqrt{\left[V(\xi) + E\right]^2 - \left[\sigma^* b(\xi - \xi_0) + V(\xi_0) + E\right]^2}\, d\xi. \tag{6}$$

For a given $V(\xi)$ the positions $\xi_0$, $\xi_c$ and the activation enthalpy $H_b$ are evaluated numerically.

## 3. Construction of the Peierls potential and the Peierls barrier

As already mentioned in the previous section, we consider that the Peierls potential, $V(x,y)$, associated with a $1/2\langle 111\rangle$ screw dislocation is a function of two variables, $x$ and $y$, that coincide with some orthogonal coordinates in the $\{111\}$ plane perpendicular to the dislocation line. This approach is similar to that advanced in [36] but the novel concept is that the *Peierls potential is considered to be a function of the full applied stress tensor*. Since atomistic molecular statics calculations do not provide full information about the Peierls potential but give only the values of the Peierls stress, $\sigma_P$, the stress dependence of the Peierls potential is extracted from the calculated dependence of the Peierls stress on the applied stress tensor (see Part I). The Peierls barrier, $V(\xi)$, is then identified with the lowest energy path, characterized by a curvilinear coordinate $\xi$, over the two-dimensional Peierls potential between two sites corresponding to equivalent minima of $V(x,y)$, i.e. between equivalent positions of the dislocation. The minimum of the Peierls barrier from which the dislocation moves is taken as $V(0) = 0$. For a given potential $V(x,y)$ the path $\xi$ is found by employing the Nudged Elastic Band (NEB) method [32, 39, 40], as explained in the Appendix. The link between the Peierls barrier and the Peierls stress, $\sigma_P$, is the relation

$$\sigma_P b = \max\left[\frac{dV(\xi)}{d\xi}\right]. \tag{7}$$

This is the fundamental relationship that allows us to make the Peierls potential dependent on the applied stress tensor when the dependence of $\sigma_P$ on this tensor is known from atomistic studies. Hence, the construction of the Peierls potential and finding the Peierls barrier using the NEB method are carried out simultaneously and self-consistently. In the remainder of this section we describe in more detail the development of the Peierls potential.

### 3.1 Symmetry mapping function

The constructed Peierls potential is based on the mapping function, $m(x,y)$, that captures the three-fold rotation symmetry associated with $\langle 111\rangle$ directions and the periodicities in $\{111\}$ planes. This function, multiplied by a constant, can be regarded as a zeroth-order approximation of the Peierls potential and we choose it to be the same as in [36], where the two-dimensional Peierls potential was first introduced. It is a product of three sinusoidal functions and for the [111] direction of the dislocation line it reads

$$m(x,y) = \frac{1}{2} + \frac{4}{3\sqrt{3}} \sin\frac{\pi}{3a_0}\left(2y\sqrt{3} + a_0\right) \sin\frac{\pi}{a_0}\left(\frac{y}{\sqrt{3}} - x - \frac{a_0}{3}\right) \sin\frac{\pi}{a_0}\left(\frac{y}{\sqrt{3}} + x + \frac{2a_0}{3}\right), \quad (8)$$

where $x$ is measured along the $[\bar{1}2\bar{1}]$ direction that is the trace of the $(\bar{1}01)$ plane in the (111) plane, and $y$ along the $[\bar{1}01]$ direction. This function is depicted as a contour plot in Fig. 2a, where dark shading corresponds to minima and light shading to maxima. It is three-fold symmetric and bounded such that $0 \leq m(x,y) \leq 1$. The minima and maxima of $m(x,y)$ form a triangular lattice with the lattice parameter $a_0 = a\sqrt{2}/3$, where $a$ is the lattice constant of the cubic lattice.

The low-energy path connecting two minima of $m(x,y)$ that we identify with the Peierls barrier always passes through the region in between neighboring minima and maxima of the Peierls potential. The sinusoidal character of the function $m(x,y)$ leads naturally to a sharp maximum when following such path. However, the derivative $dV(\xi)/d\xi$ reaches its largest value far away from this maximum and thus equation (7) that provides for the input of data from atomistic studies conveys no information about the Peierls barrier near its maximum. Auspiciously, a helpful information about this region of the Peierls barrier was obtained in [41-43] where the thermally activated formation of kink-pairs for various shapes of the Peierls barrier was investigated. The best agreement between calculated temperature dependence of the yield stress and experiments was attained when the top of the Peierls barrier exhibited a flat plateau or even an intermediate local minimum. Hence, the first modification of the function $m(x,y)$ we introduce is that the region of the Peierls potential marked by the shaded circle in Fig. 2a becomes flat while $m(x,y)$ remains unaffected elsewhere. In order to accomplish this adjustment we center at the points equivalent to the center of the shaded circle the function

$$f(r) = 1 - \frac{\beta}{1 + \exp\left(\frac{r - r_0}{\alpha}\right)}, \quad (9)$$

where $r$ is the distance from this point and $r_0$, $\alpha$ and $\beta$ are adjustable parameters. This is shown schematically in the inset of Fig. 2a. This choice is, of course, not unique but the details of the functional form producing a flat maximum along the low-energy path connecting two minima have negligible effect on activation enthalpies of the kink-pair formation. The points at which the function $f(r)$ is centered are corners of the triangular lattice in the (111) plane with primitive translation vectors $\mathbf{t}_1 = (1,0)a_0$ and $\mathbf{t}_2 = (1/2, \sqrt{3}/2)a_0$, the origin of which coincides, for example, with $\mathbf{t}_0 = (1/2, -\sqrt{3}/6)a_0$. The positions of the nodes of this lattice are determined by a set of vectors $\mathbf{t}_{k,l} = \mathbf{t}_0 + k\mathbf{t}_1 + l\mathbf{t}_2$ where $k$, $l$ are integers. Hence the modified mapping function reads

$$\tilde{m}(x,y) = F(x,y)\, m(x,y), \quad (10)$$

with

$$F(x,y) = \prod_{k,l} f(|\mathbf{r} - \mathbf{t}_{k,l}|) \tag{11}$$

where $\mathbf{r}$ is the vector with components $x,y$. Obviously, parameters $\alpha$ and $\beta$ can always be adjusted such that the function $F$ changes the mapping function significantly only in the vicinity of points defined by vectors $\mathbf{t}_{k,l}$. The extent of this change is controlled by the parameter $r_0$ and we choose $r_0 = a_0/3\sqrt{3}$ which is one-third of the distance from any such point to the nearest maximum/minimum of $m(x,y)$. The choice of parameters $\alpha$ and $\beta$ is not unique, but the mapping function should not lead to sudden drops in $\tilde{m}(x,y)$ (if $\alpha$ is too small) or intermediate local minima (if $\beta$ is too large). The values of $\alpha$ between 0.05 and 0.2, and $\beta$ between 0.1 and 0.4 assure such behavior of $\tilde{m}(x,y)$. Fig. 2b shows the contour plot of $\tilde{m}(x,y)$ calculated for $\alpha = 0.12$ and $\beta = 0.2$ that were found to be appropriate for both Mo and W (see also Table 1). The difference between $\tilde{m}(x,y)$ and $m(x,y)$ is visible by comparing this figure with Fig. 2a. The shapes of the isolines in the vicinity of the points defined by vectors $\mathbf{t}_{k,l}$ reveal that the mapping function is indeed flattened but at the same time the modified mapping function $\tilde{m}(x,y)$ still possesses the three-fold rotational and the long-range translational symmetry, as required.

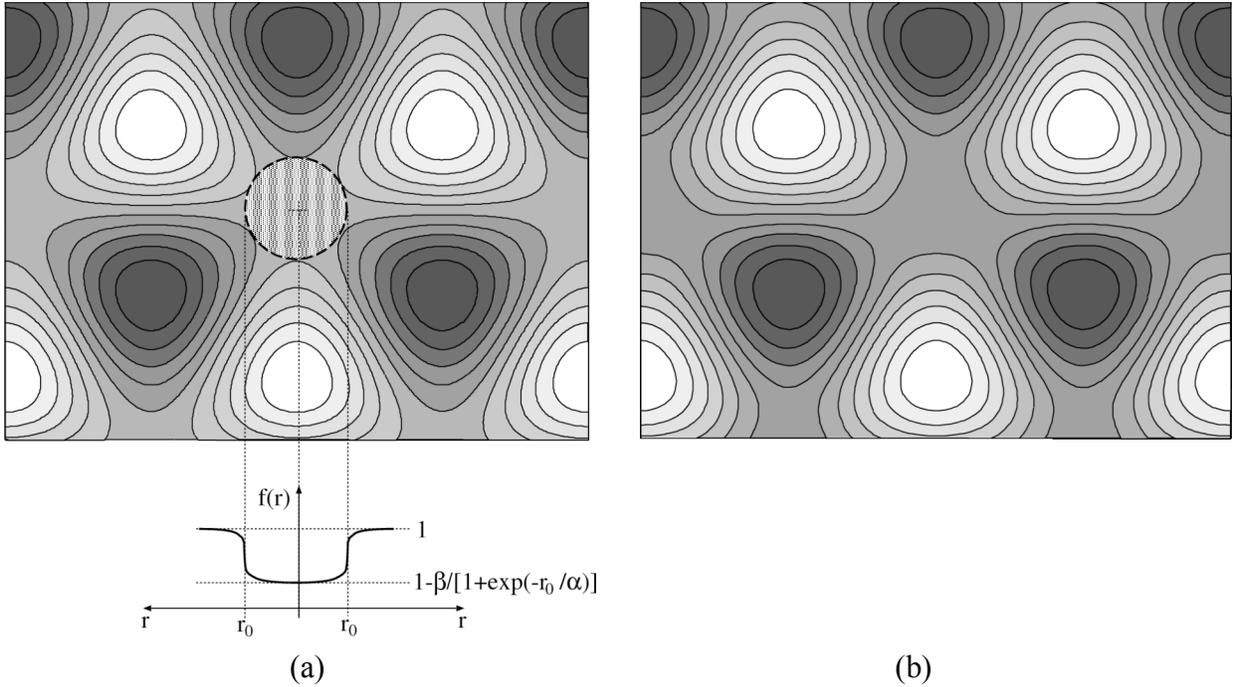

(a)          (b)

Fig. 2: (a) Contour plot of the mapping function $m(x,y)$ determined by (8). The adjustment by the function $f(r)$, given by (9), takes place primarily in the region marked by the circle of radius $r_0$. The inset shows the shape of the function $f(r)$. (b) Contour plot of the mapping function $\tilde{m}(x,y)$, determined by (10) for $\alpha = 0.12$ and $\beta = 0.2$. In both figures the dark domains correspond to minima and bright domains to maxima.

## 3.2 Peierls potential independent of applied stress

The Peierls potential that does not depend on the applied stress tensor can now be written as

$$V(x,y) = \Delta V \, \tilde{m}(x,y), \qquad (12)$$

where $\Delta V$ is the maximum height of the potential that can be determined when considering the loading by pure shear stress parallel to the [111] direction in the $(\bar{1}01)$ plane. The corresponding Peierls stress, $\sigma_P$, is equal to the CRSS for $\chi = 0$ found in atomistic simulations that are presented in Part I. $\Delta V$ can then be determined using (7) by the following self-consistent procedure. A trial value of $\Delta V$ is chosen and the NEB method used to evaluate for the Peierls potential given by (12) the minimum energy path, $\xi$, between its adjacent minima on the $(\bar{1}01)$ plane. Using the Peierls barrier $V(\xi)$ obtained in this way we evaluate $\max\left[dV(\xi)/d\xi\right]$ and compare with $\sigma_P b$. We then adjust $\Delta V$ and repeat the whole process until the difference $\max\left[dV(\xi)/d\xi\right] - \sigma_P b$ becomes less than a specified tolerance, which in our case was taken as $10^{-4}$ eV/Å$^2$. The values of $\Delta V$ found in this way for Mo and W are summarized in Table 1.

## 3.3 Dependence of the Peierls potential on the shear stress parallel to the slip direction

The atomistic study of the dependence of the CRSS on the orientation of the MRSSP when only the shear stress parallel to the slip direction is applied has been reported in Section 4.1 of Part I. With the orientation of the MRSSP characterized by angle $\chi$, defined in Fig. 2 of Part I, the CRSS for molybdenum is higher for $\chi > 0$ than for $\chi < 0$. This is the well-recognized twinning-antitwinning asymmetry common to many BCC metals [3, 5, 24, 25]. This orientation dependence of the CRSS suggests that the Peierls barrier varies with the orientation of the MRSSP such that it is higher for $\chi > 0$ and lower for $\chi < 0$. In order to account for this asymmetry, we augment the Peierls potential such that it becomes

$$V(x,y) = \left[\Delta V + V_\sigma(\chi,\theta)\right]\tilde{m}(x,y), \qquad (13)$$

where the angularly dependent function $V_\sigma(\chi,\theta)$ represents a distortion of the three-fold symmetric basis, $\tilde{m}(x,y)$, by the shear stress parallel to the slip direction. Here, $\theta$ is the angle between the x-axis and the line connecting the origin with the point $(x,y)$. A simple functional form

$$V_\sigma(\chi,\theta) = K_\sigma(\chi)\sigma b^2 \cos\theta \qquad (14)$$

proved to be sufficient to reproduce the orientation dependence of the Peierls stress found in atomistic calculations; the factor $b^2$ assures correct dimension of $V_\sigma$. $K_\sigma(\chi)$ has to be determined numerically so that the twinning-antitwinning asymmetry of the CRSS is reproduced correctly. At this point, rather than using directly the atomistic results, we make use of the yield criterion for pure shear stress parallel to the slip direction given by equation (3) of Part II:

$$\mathrm{CRSS}(\chi) = \frac{\tau^*_{cr}}{\cos\chi + a_1 \cos(\chi + \pi/3)}. \qquad (15)$$

For any orientation of the MRSSP, the dislocation glides on the $(\bar{1}01)$ plane and thus the Peierls stress entering (7) can be written as $\sigma_P = \mathrm{CRSS}(\chi)\cos\chi$. The corresponding $K_\sigma(\chi)$ can then be determined as follows. For a given $\chi$ we start with an initial guess of $K_\sigma$ and determine a trial Peierls potential following equations (13) and (14). The NEB method is then used to find the minimum energy path, $\xi$, between two adjacent potential minima on the $(\bar{1}01)$ plane. We then evaluate $\max[dV(\xi)/d\xi]$ for the Peierls barrier $V(\xi)$ obtained in this way and compare with $\sigma_p b$. $K_\sigma$ is then adjusted and we repeat the whole process until the value of $K_\sigma$ for which the Peierls stress, $\sigma_P$, is reproduced with the precision of $10^{-4}$ eV/Å$^2$ is attained

In the case of molybdenum, $K_\sigma$ is approximated well by a linear function given in Table 1. For positive $\chi$, i.e. shearing in the antitwinning sense, $V_\sigma$ is positive and both the Peierls barrier and the Peierls stress increase relative to the case of $\chi = 0$. In contrast, for negative $\chi$, i.e. shearing in the twinning sense, $V_\sigma$ is negative and both the Peierls barrier and the Peierls stress decrease relative to the case of $\chi = 0$. Consequently, the Peierls potential given by equations (13) and (14) reproduces the twinning-antitwinning asymmetry of the CRSS when loading by the shear stress parallel to the slip direction. For molybdenum the shape of the Peierls barrier $V(\xi)$ is shown in Fig. 3 for $\chi = 0$ and $\pm 20°$. Since in tungsten the CRSS exhibits virtually no twinning-antitwinning asymmetry (see Section 4.1 of Part I) $V_\sigma$ is zero in this case.

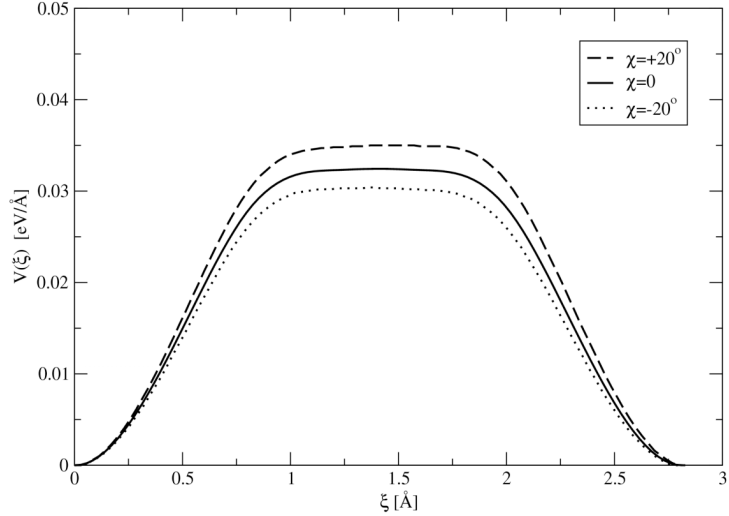

Fig. 3: The Peierls barrier, $V(\xi)$, in molybdenum corresponding to loading by pure shear stress parallel to the slip direction for three different orientations of the MRSSP. The Peierls stress that is proportional to the maximum slope of $V(\xi)$ increases with increasing angle $\chi$. This reflects the twinning-antitwinning asymmetry of shearing.

An important feature of $V_\sigma$ is that upon reversing the sense of shearing, $V_\sigma(\chi,\theta)$, given by (14), changes sign and since $-K_\sigma(\chi) = K_\sigma(-\chi)$ this is equivalent to keeping the sign of the shear stress the same and reversing the sign of the angle $\chi$. This is the well-known feature of the twinning-antitwinning asymmetry and it is correctly reproduced by the $V_\sigma$ term. Finally, it is important to note that $V_\sigma$ breaks the translational symmetry of the Peierls potential. Consequently, when employing $V(x,y)$ given by (13) in a study of dislocation motion the origin of $V(x,y)$ must always coincide with the site from which the dislocation moves into the adjacent site. Hence, after each elementary step is completed, the origin of $V(x,y)$ has to be shifted to this new position.

### 3.4 Dependence of the Peierls potential on the shear stress perpendicular to the slip direction

In order to incorporate the effect of the shear stress $\tau$, perpendicular to the slip direction, we supplement $V(x,y)$ by the term $V_\tau(\chi,\theta)$ that represents the distortion of the Peierls potential by $\tau$. The meaning of angles $\chi$ and $\theta$ is the same as in the previous section. The Peierls potential that comprises the effects of both the shear stress parallel to the slip direction in a plane other than the slip plane and the shear stress perpendicular to the slip direction is then

$$V(x,y) = \left[\Delta V + V_\sigma(\chi,\theta) + V_\tau(\chi,\theta)\right]\tilde{m}(x,y). \tag{16}$$

To be consistent with the yield criterion given by Eq. (2) of Part II that depends linearly on $\tau$, we require that also $V_\tau$ is a linear function of $\tau$. In addition, since the stress tensor composed of the shear stresses parallel and perpendicular to the slip direction, given by Eq. (2) of Part II, is invariant with respect to rotations by integral multiples of $\pi$ around the $z$-axis, $V_\tau$ has to obey the same symmetry. The simplest form of $V_\tau$ that satisfies these requirements is

$$V_\tau(\chi,\theta) = K_\tau(\chi)\tau b^2 \cos(2\theta + \pi/3), \tag{17}$$

where $K_\tau(\chi)$ is to be determined numerically by fitting atomistically calculated dependence of the Peierls stress on $\chi$ and $\tau$ and the factor $b^2$ again assures correct dimension of $V_\tau$. When determining $K_\tau(\chi)$ we consider, similarly as in Section 3 of Part II, only those values of $\tau$ for which $|\tau/C_{44}| \leq 0.02$ [1]. For these stresses, the dislocation glides on the $(\bar{1}01)$ plane in both molybdenum and tungsten and thus the Peierls stress in (7) can be again written as $\sigma_P = \text{CRSS}(\chi)\cos\chi$. The CRSS can now be determined using the yield criterion for the combined shear stresses parallel and perpendicular to the slip direction given by Eq. (4) of Part II:

$$\text{CRSS}(\chi,\tau) = \frac{\tau_{cr}^* - \tau[a_2 \sin 2\chi + a_3 \cos(2\chi + \pi/6)]}{\cos\chi + a_1 \cos(\chi + \pi/3)}. \tag{18}$$

The benefit of this approach is that we do not consider a priori that the dislocation may glide in a different {110} plane as a result of the transformation of its core induced by $\tau$. Rather, if the Peierls potential is constructed correctly, this change of the glide plane will be predicted. This provides an important test of the predictive power of the constructed Peierls potential. Such change of the slip plane driven by $\tau$ is, indeed, seen in Fig. 4 that is discussed below.

$K_\tau(\chi)$ can be determined for a given $\chi$ similarly as $K_\sigma(\chi)$ in Section 3.3. We start again with an initial guess of $K_\tau$ and determine a trial Peierls potential following equations (16) and (17); the values of $\Delta V$ and $K_\sigma(\chi)$ entering (16) are those determined as described in Sections 3.2 and 3.3. The minimum energy path, $\xi$, between two adjacent potential minima on the $(\bar{1}01)$ plane is then determined using the NEB method. For this path $\max\left[dV(\xi)/d\xi\right]$ is evaluated, compared with $\sigma_P b$ and $K_\tau$ adjusted accordingly. The whole process is then repeated until the value of $K_\tau$ for which the Peierls stress $\sigma_P$ is reproduced with the precision of $10^{-4}$ eV/Å$^2$ is attained. For both molybdenum and tungsten $K_\tau(\chi)$ can be closely approximated by a quadratic polynomial presented in Table 1.

---

[1] As shown in the Part II, larger values of $\tau$ are virtually inaccessible in real single crystals since another {110}⟨111⟩ system becomes operative.

Table 1: Parameters of the Peierls potential (16) for molybdenum and tungsten. The parameters α and β entering the function f(r) given by (9) are the same for Mo and W. The angle χ is measured in radians.

| | $a_0$ | α | β | $\Delta V$ [eV/Å] | $K_\sigma(\chi) = k\chi$ | $K_\tau(\chi) = c_0 + c_1\chi + c_2\chi^2$ | | |
|---|---|---|---|---|---|---|---|---|
| | | | | | $k$ | $c_0$ | $c_1$ | $c_2$ |
| Mo | 2.570 | 0.12 | 0.2 | 0.0787 | 0.139 | -0.171 | 0.182 | 0.319 |
| W | 2.584 | | | 0.1369 | 0 | -0.413 | -0.216 | 0.782 |

The effect of the pure shear stress perpendicular to the slip direction on the shape of the Peierls potential is shown in Fig. 4 where the potential is displayed as a contour plot for three different values of τ. Corresponding minimum energy paths between adjacent potential minima, determined by the NEB method, are shown as curves. It is evident that positive shear stress perpendicular to the slip direction lowers the potential barrier for the slip on the $(\bar{1}01)$ plane and, therefore, makes the slip on this plane easier. In contrast, for negative τ the Peierls barrier for the $(\bar{1}01)$ slip increases while it decreases for the slip along $(\bar{1}10)$ and $(0\bar{1}1)$ planes owing to the flattening of the potential along these planes. Hence, for negative τ the slip on either $(0\bar{1}1)$ or $(\bar{1}10)$ plane is more likely. This agrees with the findings of atomistic simulations shown in Figs. 7 and 8 of Part I, which demonstrates that the predictions based on the Peierls potential (16) are fully consistent with the results of atomistic calculations at 0 K. Furthermore, $|K_\tau(\chi)|$ for tungsten is always larger than for molybdenum and this means that the effect of the shear stress perpendicular to the slip direction is appreciably stronger in tungsten than in molybdenum, as seen in atomistic studies presented in Part I.

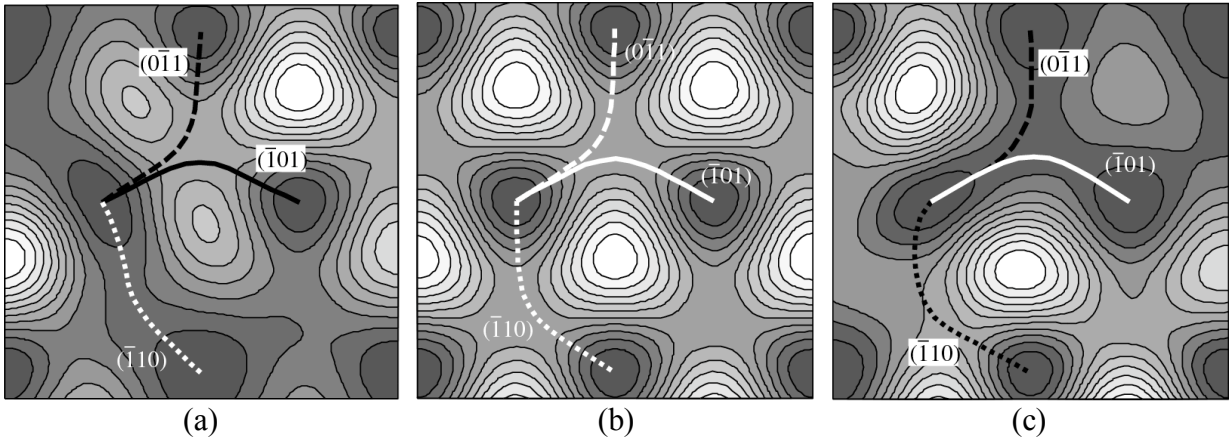

(a)    (b)    (c)

Fig. 4: Contour plot of the Peierls potential given by (16) for three different values of the applied shear stress τ perpendicular to the slip direction (no shear stress parallel to the slip direction is applied): (a) $\tau/C_{44} = -0.05$, (b) $\tau/C_{44} = 0$, (c) $\tau/C_{44} = +0.05$. Minimum energy paths between adjacent potential minima are shown as curves. The activation paths characterized by the low Peierls barriers are drawn in white. Note that in (c) the low-energy path of the dislocation lies along the $(\bar{1}01)$ plane that is the {110} plane with the highest resolved shear stress in the slip direction when $-30° \leq \chi \leq +30°$. In contrast, in (a) the low-energy path is along the $(\bar{1}10)$ plane in which the resolved shear stress in the slip direction is always lower for this range of χ.

## 4. Activation enthalpy for dislocation glide and temperature dependence of the yield stress

Within the model of kink-pair formation, described in Section 2, the activation enthalpy, $H(\sigma)$, for thermally activated dislocation glide is for low stresses and high temperatures equal to $H_{kp}$, given by (3b), and at high stresses and low temperatures to $H_b$, given by (6). The transition between the two stress regimes occurs for the stress at which $H_{kp} = H_b$. Consequently, at low stresses the activation enthalpy is for a given $\{110\}\langle111\rangle$ slip system only a function of the shear stress in this $\{110\}$ plane applied parallel to the slip direction, specifically its part $\sigma^*$ defined by (2). Other components of the applied stress tensor have in this case no influence. In contrast, at high stresses $H(\sigma)$ is for a given $\{110\}\langle111\rangle$ slip system a function of the shear stress parallel to the slip direction in the MRSSP and orientation of this plane relative to the corresponding $\{110\}$ plane, defined by the angle $\chi$, as well as a function of the shear stress perpendicular to the slip direction. This activation enthalpy, evaluated according to (6), is determined by the stress-dependent Peierls barrier introduced in the previous section. However, in both cases the dependence of the yield stress on temperature and strain rate is for a given $\{110\}\langle111\rangle$ slip system determined by $\sigma^*$ that is the part of the shear stress parallel to the slip direction in the corresponding $\{110\}$ plane.

When evaluating the activation enthalpy for low stresses, $H_{kp}$, we need to know the energy of an isolated kink $H_k$. This energy could be calculated atomistically as it was done by Duesbery [37, 38] using pair potentials. However, since for $\sigma^* = 0$ the activation enthalpy is equal to $2H_k$, the energy of a kink can be estimated with high precision from experimental data of the temperature dependence of the flow stress at low stresses [44, 45] and/or from studies of internal friction [46]. Values of $2H_k$ determined in this way for Mo and W are presented in Table 2. When evaluating $H_b$ using (6) the value of the line tension, E, needs to be fixed and in the present calculations it was taken as $\mu b^2 / 4$, where $b$ is the magnitude of the Burgers vector and $\mu$ the $\{110\}\langle111\rangle$ shear modulus $(C_{11} - C_{12} + C_{44})/3$; its values for Mo and W are given in Table 2.

The model of the dislocation motion at finite temperatures that involves overcoming the Peierls barrier with the aid of thermal activation has to be consistent with the movement of the 1/2[111] screw dislocation at 0 K investigated in atomistic studies. Such consistency is attained if the activation enthalpy becomes zero when $\sigma^*$ is equal to the Peierls stress. It is then the stress at which $\xi_0 = \xi_c$ and, therefore, equation (5) is automatically satisfied. However, all atomistic studies of the glide of screw dislocations in BCC metals give the Peierls stress that is by a factor of two to three larger than the CRSS obtained by extrapolating low-temperature experimental measurements of the yield and flow stresses to 0 K. This problem was thoroughly documented by comparison of calculated and experimental data in [47] and explained on the basis that in reality dislocations never move in isolation but are produced by sources and large groups consisting of non-screw dislocations with low Peierls stress near the source and screw dislocations further away. This is in contrast with atomistic modeling of a single screw dislocation. Owing to the mutual interaction between dislocations emitted from a source the stress acting on the screw dislocations positioned at a distance from this source is about a factor of two to three higher than the applied stress. Hence, when comparing calculations with experiments all the calculated stresses have to be scaled such that $\sigma^*$ at 0 K is equal to the yield stress found by extrapolating experimental measurements to 0 K.

An example of the dependence of the activation enthalpy on the shear stress $\sigma^*$ is shown in Fig. 5 for the $(\bar{1}01)[111]$ system in the single crystal of molybdenum loaded in tension/compression along the $[\bar{1}49]$ axis; in this case the MRSSP is the $(\bar{1}01)$ plane and $\chi = 0$. For this loading comparison with experimentally observed temperature dependence of the yield stress [44, 48] is presented in Section 5. The above-mentioned scaling is congruent with dividing the calculated stress by the factor of 2.9 and the values of $\sigma^*$ in Fig. 5 have been scaled in this way. It is seen that the model predicts a significant tension-compression asymmetry since for $\sigma^* > 150$ MPa the activation enthalpy for compression is higher than that for tension. This asymmetry, which is compatible with the results of atomistic studies presented in Part I, increases with increasing $\sigma^*$ and thus decreasing temperature. However, it disappears at stresses below about 150 MPa (and thus high temperatures) when the saddle-point configuration for the formation of kink-pairs changes from the bow-out to a pair of fully formed kinks.

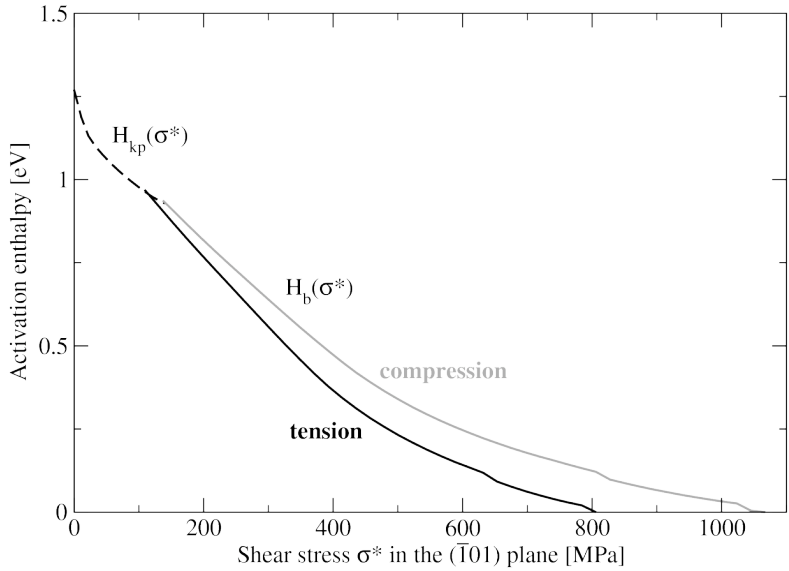

Fig. 5: Dependence of the activation enthalpy for the $(\bar{1}01)[111]$ system in molybdenum on the temperature dependent component of the applied shear stress, $\sigma^*$, for loading in tension/compression along the $[\bar{1}49]$ axis for which $\chi = 0$. Full curves correspond to the activation enthalpy given by (6) (bow-out) and dashed curve to the activation enthalpy given by (3b) (fully developed kinks).

The temperature dependence of the yield stress $\sigma^*$ for a given slip system at a fixed plastic strain rate $\dot{\gamma}$ can be determined from equation (1b) when written as

$$H(\boldsymbol{\sigma}) = k_B T \ln(\dot{\gamma}_0 / \dot{\gamma}). \tag{19}$$

By solving[2] (19) for $\sigma^*$, in general numerically, we obtain $\sigma^*(T, \dot{\gamma})$. The pre-exponential factor $\dot{\gamma}_0$ can be written approximately as [9, 14, 15] $\dot{\gamma}_0 \approx b^2 a_0 \rho_m \upsilon_D / \ell$, where b is the magnitude of the Burgers vector of the moving dislocation, $a_0$ the distance the dislocation moves during one activation step, i.e. the separation between two neighboring Peierls valleys, $\rho_m$ the mobile dislocation density in the slip system considered, $\upsilon_D$ the Debye frequency and $\ell$ the lateral extent of the kink-pair in the saddle-point configuration. However, the mobile dislocation density is generally not known and thus this evaluation of $\dot{\gamma}_0$ introduces large uncertainty into determination of $\sigma^*(T, \dot{\gamma})$. Notwithstanding, when comparing calculations with experiments that have usually been carried out at a fixed plastic strain rate $\dot{\gamma}$, the factor $\dot{\gamma}_0$ can be determined from the temperature $T_k$ at which the thermal component of the yield stress vanishes. At this temperature the activation enthalpy is equal to $2H_k$ and thus from (19) we obtain

$$\ln(\dot{\gamma}_0/\dot{\gamma}) = 2H_k/k_B T_k .  \qquad (20)$$

---

[2] Note that $\sigma^*$ is a part of the full stress tensor $\boldsymbol{\sigma}$.

The values of the temperature $T_k$ and the corresponding values of $\ln(\dot{\gamma}_0/\dot{\gamma})$ obtained for Mo and W from experiments in [44] and [45], respectively, are given in Table 2.

Table 2: Energy of two isolated kinks deduced from experiments in [44] and [45], the shear modulus μ entering the expression for the line tension, temperature $T_k$ at which the thermal component of the yield stress in tension vanishes and corresponding values of $\ln(\dot{\gamma}_0/\dot{\gamma})$.

|    | $2H_k$ [eV] | μ [eV/Å³] | $T_k$ [K] | $\ln(\dot{\gamma}_0/\dot{\gamma})$ |
|----|-------------|-----------|-----------|-------------------------------------|
| Mo | 1.27        | 0.789     | 472       | 31.2                                |
| W  | 2.06        | 1.143     | 760       | 31.4                                |

Finally, it should be noted that several slip systems may operate concurrently when deforming a single crystal and thus the total plastic strain is

$$\dot{\gamma}_{tot} = \sum_{\alpha} \dot{\gamma}_0^{\alpha} \exp\left[-H^{\alpha}(\boldsymbol{\sigma})/k_B T\right],  \qquad (21)$$

where the summation extends over all active slip systems. However, owing to the exponential dependence on the activation enthalpy, just a small number of slip systems, and frequently only one, contribute significantly to the plastic strain rate.

## 5. Comparisons with experimental measurements of the temperature dependence of the yield stress

The temperature dependence of the yield stress in molybdenum single crystals loaded in tension was measured in [44, 48, 49] for three different orientations of the tensile axis and we compare here our calculations with measurements for the loading along the $[\bar{1}49]$ axis, when χ = 0. In these experiments the plastic strain rate is $\dot{\gamma} = 8.6 \times 10^{-4}$ s⁻¹ and the temperature at which the thermal component of the yield stress vanishes was estimated as $T_k = 472 K$. Using the relation (20) we obtain the value of $\ln(\dot{\gamma}_0/\dot{\gamma})$ given in Table 2 that corresponds to $\dot{\gamma}_0 = 3 \times 10^{10}$ s⁻¹. This value suggests the mobile dislocation density about $10^{17}$ m⁻² if considering, as construed above, that $\dot{\gamma}_0 \approx b^2 a_0 \rho_m \upsilon_D / \ell$. This is a reasonable value for plastically deformed crystals [9, 11]. Using the values summarized in Tables 1 and 2, equation (19) was solved for $\sigma^*$ for both tension and compression. The results of this calculation are presented in Fig. 6 together with the experimental data for tension from [48]. As explained in the previous section, the values of $\sigma^*$ plotted as solid curves in Fig. 6 correspond to the calculated ones divided by 2.9, similarly as in Fig. 5, so that $\sigma^*$ at 0 K is equal to the yield stress found by extrapolating experimental measurements to 0 K.

At temperatures below 350 K, where the bow-out mechanism of kink-pair formation operates, $\sigma^*$ is higher in compression than in tension. This is in agreement with atomistic studies presented in Part I and with experimental observations of the tension-compression asymmetry in [49]. This asymmetry is best described by the strength differential *SD* that was defined by Eq. (6) of Part II in terms of the uniaxial yield stresses $\sigma_t$ and $\sigma_c$ for tension and compression, respectively. It is important to emphasize that $\sigma^*$ varies with temperature in the same way for both tension and compression and thus the same applies to $\sigma_t$ and $\sigma_c$. Therefore, *SD* is practically temperature-independent. Comparison between calculated values of *SD* and those determined experimentally in [49] is presented in Table 3 of Part II. It is particularly interesting that $SD < 0$ when the $(\bar{1}01)$ plane is the MRSSP ($\chi = 0$). If the twinning-antitwinning asymmetry were the sole reason for the tension-compression asymmetry then no asymmetry would be observed for this orientation of the MRSSP. In the framework of our model the reason for this asymmetry is the effect of shear stresses perpendicular to the slip direction, as was shown in Section 7 of Part II.

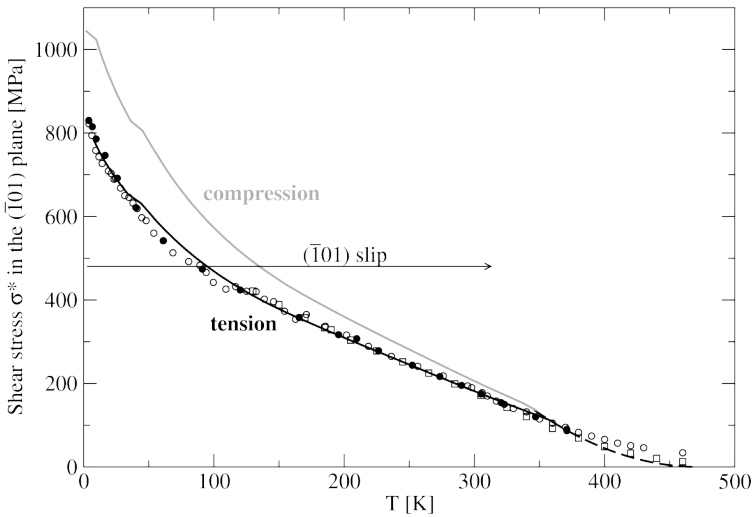

Fig. 6: Temperature dependence of the yield stress for the $(\bar{1}01)[111]$ system in molybdenum calculated for loading in tension/compression along $[\bar{1}49]$ (curves) using the parameters summarized in Table 2. The calculated values of $\sigma^*$ are divided by 2.9 as explained in Section 4. The experimental data (symbols) for tension and strain rate $\dot{\gamma} = 8.6 \times 10^{-4}$ s$^{-1}$ obtained using several different methods are from [48].

A similar plot of the dependence of $\sigma^*$ on temperature for tungsten deformed in tension/compression along the $[\bar{1}49]$ axis is presented in Fig. 7. The calculations are compared with the experimental data obtained at the strain rate of $\dot{\gamma} = 8.5 \times 10^{-4}$ s$^{-1}$ in [45]. The parameters needed for the calculation of $\sigma^*$ are summarized in Table 2 and the values of $\sigma^*$ plotted as solid curves in Fig. 7 correspond to the calculated ones divided by the factor of 3.7 so that $\sigma^*$ at 0 K is the same as the yield stress found by extrapolating experimental measurements to 0 K. Similarly as in Mo, at temperatures below about 650 K, where the bow-out mechanism of kink-pair formation operates, $\sigma^*$ is higher in compression than in tension and this asymmetry is larger than in molybdenum. This finding is in agreement with atomistic studies presented in Part I. An interesting result of the calculations is that while the $(\bar{1}01)[111]$ system operates at all temperatures in tension, in compression the system $(\bar{1}10)[\bar{1}\bar{1}\bar{1}]$ becomes dominant at temperatures below 150 K. Unfortunately, no experimental measurements of the yield stress in compression that would be of similar accuracy as the measurements in [45] are available to test these predicted features of the plastic yielding in tungsten.

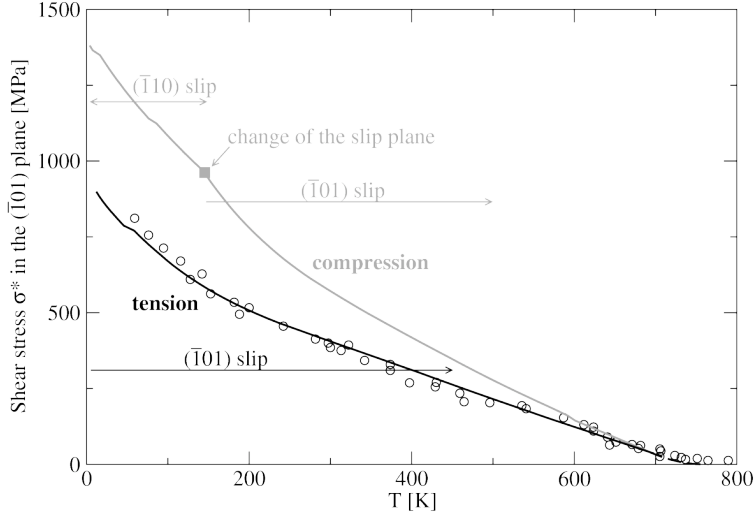

Fig. 7: Temperature dependence of the yield stress for the $(\bar{1}01)[111]$ system in tungsten calculated for loading in tension/compression along $[\bar{1}49]$ (curves) using the parameters summarized in Table 2. The calculated values of $\sigma^*$ are divided by 3.7 as explained in Section 4. The experimental data (circles) for tension and strain rate $\dot{\gamma}=8.5\times10^{-4}$ s$^{-1}$ are from [45].

Fig. 8 shows the activation volume, defined as $v^* = -\partial H(\sigma)/\partial\sigma^*$, calculated for glide on the $(\bar{1}01)[111]$ system when the crystal of molybdenum is loaded in tension along the $[\bar{1}49]$ axis[2]. Following (1b) the activation volume relates to the strain rate sensitivity such that $v^* = k_B T \left[ \partial \ln(\dot{\gamma}/\dot{\gamma}_0)/\partial\sigma^* \right]_T$ and this relation has been commonly used to measure the activation volume. The activation volume in molybdenum single crystals loaded in tension was measured in [50, 51]. For the case of $\chi = 0$, albeit not the $[\bar{1}49]$ tensile axis, the experimentally determined values of $v^*$ are plotted in Fig. 8 as circles. The agreement between calculated and measured activation volumes is qualitatively very good. For $\sigma^* > 200$ MPa the activation volume, $v^*$, varies with $\sigma^*$ only slowly, between a few b$^3$ up to about 20b$^3$. However, $v^*$ increases rapidly for small $\sigma^*$. Within our model the regime of the slow increase of $v^*$ corresponds to the saddle-point configuration being a dislocation bow-out while a rapid increase of $v^*$ ensues when the saddle-point configuration corresponds to a well-formed pair of kinks. This form of the dependence of $v^*$ on $\sigma^*$, as well as its magnitude, was observed in measurements of $v^*$ for a variety of single and polycrystalline BCC metals [50, 52, 53]. Hence, it is characteristic for the thermally activated overcoming of the Peierls barrier related to the non-planar cores of screw dislocations.

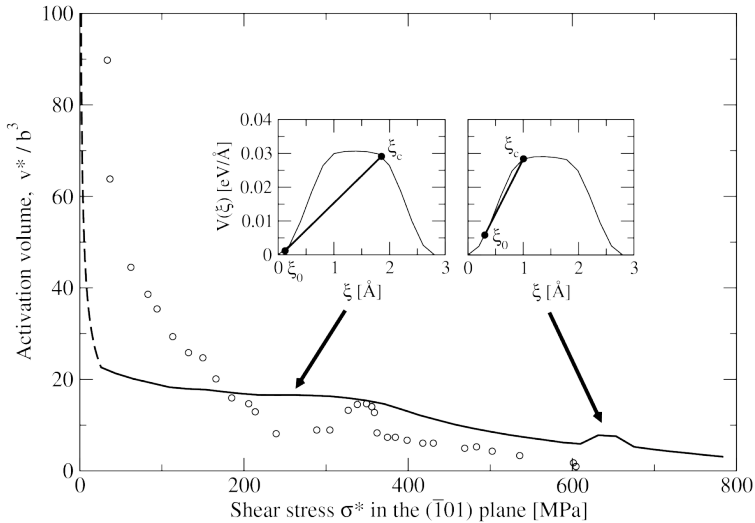

Fig. 8: Stress dependence of the activation volume for Mo loaded in tension along the $[\bar{1}49]$ axis when the plastic deformation is confined to the $(\bar{1}01)[111]$ system with the $(\bar{1}01)$ plane as the MRSSP. The experimental data of Aono et al. [51], depicted as circles, correspond to the same type of loading. The insets explain the occurrence of humps.

An interesting feature of the stress dependence of the activation volume is the occurrence of humps at certain stress levels. In the experimental data [51] shown in Fig. 8 it occurs around 350 MPa and in the calculations at the same stress level, albeit in a less pronounced way, and again around 650 MPa. This phenomenon originates in the flat plateau at the maximum of the Peierls barrier as shown in the insets of Fig. 8. Around the stresses 350 MPa and 650 MPa, the positions of $\xi_c$ corresponds to the points at which the derivative of $V(\xi)$ varies rapidly owing to the transition from the region of fast increase (decrease) of $V(\xi)$ to the region where $V(\xi)$ is almost constant. This variation is then reflected in the rapid change of the derivative of the activation enthalpy and this leads to the hump in the activation volume. This correlation between observations and calculations provides a further support for the construction of the Peierls potential described in Section 3, in particular for introduction of a flat plateau at the maximum of the Peierls barrier.

## 6. Discussion and conclusions

The principal aim of this paper has been to develop a link between the insights into the glide of $1/2\langle 111\rangle$ screw dislocations in BCC metals, specifically Mo and W, gained by atomic-level modeling at 0 K and the thermally activated glide of these dislocations at finite temperatures that proceeds via nucleation and subsequent propagation of pairs of kinks. Traditionally, the formation of kink-pairs was considered as a mechanism for overcoming the periodically varying Peierls barrier that is a stress-independent property of the material. In contrast, in this paper the Peierls barrier is regarded as dependent on the applied stress tensor, in particular shear stresses parallel and perpendicular to the slip direction. This dependence has the same origin and is of the same type as that found for the Peierls stress in 0 K atomistic studies of the dislocation glide. Moreover, in the conventional way of thinking the dislocation moves under the effect of the applied shear stress acting in the slip plane in the direction of the Burgers vector continually through the lattice, up the Peierls barrier, to metastable positions with higher energy. The Peierls stress is then the shear stress at which no metastable position exists. However, the atomistic calculations show (see Part I) that screw dislocations in BCC metals do not move continuously through the lattice but their cores gradually transform under the effect of applied stresses. The Peierls stress is then the stress at which the core structure is sufficiently altered for the dislocation to start moving. Similar complex transformations of the core structure can be expected to take place when forming the kink-pairs. However, atomistic studies of this process, for example by employing molecular dynamics, are only feasible for special cases but are impractical to obtain the complete dependence on the applied stress tensor.

In this paper we have developed a simpler phenomenological approach in which the complex core behavior is projected onto a hypothetical periodically varying Peierls barrier that reproduces all the aspects of dislocation glide at 0 K resulting from the stress-induced core transformations, in particular the dependence of the Peierls stress on the orientation of the MRSSP of shearing in the slip direction and shear stresses perpendicular to the slip direction. The dislocation is then considered as moving continually up this barrier under the effect of an applied stress. This Peierls barrier is identified with the lowest energy path over the two-dimensional Peierls potential between two equivalent locations of the dislocation. This potential is, similarly as in [36], a function of the position of the intersection of the dislocation line with the {111} plane perpendicular to the corresponding ⟨111⟩ slip direction. In an unstressed crystal the Peierls potential possesses the three-fold rotation symmetry associated with the ⟨111⟩ direction as well as the periodicity of the corresponding {111} plane. However, in the crystal subjected to loading the Peierls potential is distorted and looses the three-fold rotation symmetry but it retains its periodicity. It is the distortion of the Peierls potential that is contrived such as to reproduce the dependence of the Peierls stress on the orientation of the MRSSP and on the shear stress perpendicular to the slip direction. This development is described in detail in Section 3. Furthermore, the region of the potential where the Peierls barrier reaches its maximum, which is not sampled in 0 K calculations, is chosen to be flat as suggested by the analysis of the effect of this region on the temperature dependence of the yield stress in [41-43].

When the Peierls barrier is known the thermally activated dislocation motion via formation of kink-pairs is treated in this paper using standard dislocation models and the temperature dependence of the yield stress is determined for a given plastic strain from (19). As seen in Figs. 6 and 7, the temperature dependence of the yield stress is described very successfully for both Mo and W. Moreover, the calculated stress dependence of the activation volume for Mo, shown in Fig. 8, displays all the features found in experimentally measured strain-rate sensitivity. In particular, the activation volume varies only slowly in the regime of high stresses (low temperatures) but increases precipitously at low stresses. The latter increase is associated with the transition from the bow-out saddle-point configuration to the fully developed kink-pair. In addition, the flat top of the Peierls barrier leads to the occurrence of humps in the stress dependence of the activation volume at some stress levels, the feature observed in several experiments [50, 51].

At low stresses and high temperatures the saddle-point configuration corresponds to a well-developed pair of kinks (Fig. 1a) and in this model, originally proposed by Seeger [1], the activation enthalpy does not depend on the form of the Peierls barrier. Hence, it depends only on the shear stress parallel to the slip direction, specifically its temperature-dependent component $\sigma^*$, defined by (2). Consequently, neither twinning-antitwinning asymmetry nor the asymmetry induced by the shear stress perpendicular to the slip direction appears in this regime of stresses and temperatures. However, the situation is entirely different at high stresses when the bow-out model, originally advanced in [2], applies (Fig. 1b). In this case the activation enthalpy depends sensitively on the Peierls barrier, and thus not only on the shear stress parallel to the slip direction but also on the related orientation of the MRSSP as well as on the shear stress perpendicular to the slip direction. The latter dependence induces the tension-compression asymmetry even when the MRSSP coincides with the most highly stressed {110} plane, and the sense of shearing (twining-antitwinning asymmetry) plays no role. For Mo this is in excellent agreement with experimental observations, as discussed in Section 5 and Part II. The tension-compression asymmetry predicted for W is even larger than for Mo. Furthermore, in compression the system $(\bar{1}10)[\bar{1}\bar{1}1]$ dominates over the most highly stressed system $(\bar{1}01)[\bar{1}\bar{1}\bar{1}]$ system at temperatures below 150 K. This is reminiscent of the *anomalous slip* that occurs on the slip systems for which the resolved shear stress is significantly lower than for the most highly stressed {110}⟨111⟩ system and that was observed in a number of BCC transition metals at low temperatures (see, for example [54-60]). Unfortunately, to our knowledge, no experimental studies of W that could be compared with these predictions have been made.

The analyses presented in the series of these three papers dealing with the plastic deformation of molybdenum and tungsten aims at the overarching goal of multiscale modeling of the deformation behavior of BCC transition metals, starting with their electronic structure and proceeding up to constitutive relations applicable in continuum mechanics, as alluded to in [61]. The very first step is to coarse-grain the problem in that the electronic degrees of freedom are removed by introducing potentials representing the interaction between the atoms that is mediated by electrons [62-65]. This has been done by employing the Bond Order Potentials [66, 67] in the atomistic studies of dislocations and their glide in the Part I of this series. These results are then used in Part II to formulate a general yield criterion for single crystals that incorporates the complex dependence of the Peierls stress on the orientation of the MRSSP and shear stresses perpendicular to the slip direction, found in atomistic studies. In this paper, Part III, the results of atomistic studies and the constructed yield criterion are employed to develop a mesoscopic model of thermally activated motion of screw dislocations that leads to the calculation of the temperature dependence of dislocation velocity and thus the temperature and strain rate dependence of the yield stress. The crucial step in this development is a projection of the dependencies of the Peierls stress on the applied stress tensor found in 0 K atomistic calculation, arising from complex core transformations, onto a stress-dependent Peierls potential that reproduces them. The next step of this development, which will be published elsewhere, is the formulation of the yield criterion that includes not only the effect of the orientation of the MRSSP and shear stresses perpendicular to the slip direction but also the effects of temperature and strain rate [68].

**Acknowledgements**

This research was supported by the Department of Energy, BES Grant no. DE-PG02-98ER45702.

**APPENDIX: Determination of the minimum energy path by the Nudged Elastic Band Method (NEB)**

The NEB method [32, 39, 40] has been used in this study to determine the Peierls barrier, $V(\xi)$, that is identified with the minimum energy path (MEP) that a dislocation takes between two minimum energy sites of a known two-dimensional Peierls potential $V(x,y)$, defined in Section 3. Within this method, one works with replicas of the system that are connected together with "springs" to obtain a discrete representation of the sought path. In the terminology of the NEB method, the individual replicas of the system are called "images". The string of $N+2$ images is represented by a chain of states $[\xi_0, \xi_1, \ldots, \xi_{N+1}]$. Two of these images, namely 0 and $N+1$, are fixed in the two low-energy configurations that represent the initial and target state of the system and the positions of the remaining $N$ images, forming the so-called elastic band, are adjusted by the optimization algorithm. In our case, $\xi_0$ is the position of the dislocation in the initial configuration and $\xi_{N+1}$ in the site into which it moved via formation and extension of the kink-pairs. The most straightforward way of obtaining the coordinates of the $N$ intermediate images is by connecting the nearest neighbors with identical linear springs, characterized by the spring constant $\kappa$, and subsequently minimizing the total potential energy represented by the objective function

$$E(\xi_1, \ldots, \xi_N) = \sum_{i=1}^{N} V(\xi_i) + \sum_{i=1}^{N+1} \frac{1}{2} \kappa (\xi_i - \xi_{i-1})^2 \quad (A1)$$

with respect to the positions of images $\xi_1, \ldots, \xi_N$. As pointed out by Henkelman et al. [40], the method as formulated is not always well-behaved in that the elastic band tends to straighten in the regions where the MEP is curved, and the images tend to slide down toward the fixed endpoints 0 and $N+1$. This results in a poor resolution of the path close to the saddle-point of the potential $V(x,y)$, i.e. the maximum of $V(\xi)$. This straightening, or corner-cutting, is caused by the component of the spring force perpendicular to the elastic band, while the down-sliding is caused by the parallel component of the so-called "true force" arising from the potential, in our case from the Peierls potential. The idea to avoid these problems is referred to as "nudging" in which each image is subjected only to the component of the spring force parallel to the elastic band and the perpendicular component of the true force. If we denote the unit vector tangent to the elastic band at image $i$ as $\hat{\tau}_i$, the force on each image is

$$\mathbf{F}_i = \mathbf{F}^s_{i\parallel} - \nabla V(\xi_i)_\perp \quad (A2)$$

where $\mathbf{F}^s_{i\parallel} = (\mathbf{F}^s_i \cdot \hat{\tau}_i)\hat{\tau}_i$ denotes the parallel component of the spring force at image $i$, and $-\nabla V(\xi_i)_\perp$ is the positive perpendicular component of the true force arising from the Peierls potential $V(x,y)$. During the minimization, the parallel component of the spring force does not interfere with the perpendicular component of the true force and $\nabla V(\xi_i)_\perp \to 0$ ($i = 1, \ldots, N$) as the shape of the elastic band approaches the MEP. In the relaxed configuration, the force on each image coincides with the parallel component of the spring force, which then determines the spacing between individual images along the path.

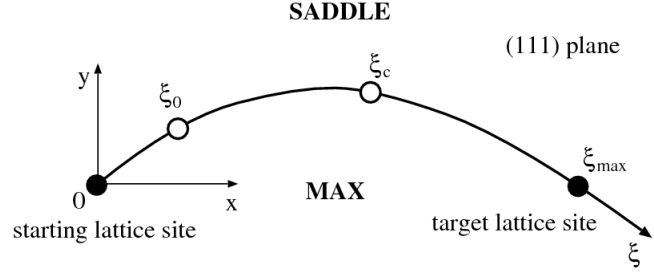

Fig. A1: Schematic representation of a curved transition coordinate $\xi$ for a dislocation moving between two minimum energy sites (filled circles). SADDLE and MAX designate a saddle-point and a maximum of the Peierls potential, respectively. At zero applied stress the kink develops from 0 to $\xi_{max}$, whereas at finite stresses the kink only extends from $\xi_0$ to $\xi_c$.

    The NEB method has been used extensively in the present work to find the MEPs of the dislocation between two minimum energy configurations and calculate the Peierls barriers along these paths. All searches utilize the method of variable spring constants with improved definition of tangents [40] in combination with the switching function defined in [33]. The method is rather insensitive to the chosen range of magnitudes of the spring constants and performed efficiently for all shapes of the Peierls potential. The velocity Verlet algorithm [69] was used to update the positions of images 1 to *N* in each iteration step.